\documentstyle[12pt]{article}
\begin{document}
\vskip 72pt
\centerline{\bf EFFICIENCY OF NEUTRON DETECTION OF} 
\centerline{\bf SUPERHEATED DROPS OF FREON-22} 

\vskip 36pt
\centerline{Mala Das, B. Roy, B. K. Chatterjee and S. C. Roy}
\centerline{\it Department of Physics, Bose Institute}
\centerline{\it 93 / 1 A. P. C. Road, Calcutta 700009, India}
\vskip 36pt
\begin{abstract}

     Neutron detection efficiency of superheated drops of  Freon-22 
for neutrons obtained from a 3 curie Am-Be neutron source has  been 
reported in this paper.  Although  Freon-22  having  lower  boiling 
point than many other  similar  liquids  (e.g.Freon-12,  Freon-114, 
Isobutane ) is expected to be more sensitive to neutrons  ,it   has 
not been reported so far and therefore this paper  constitutes  the 
first report on the subject. Neutron detection efficiency  of  both 
Freon-22 and  Freon-12  have  been  determined  from  the  measured 
nucleation rate  using  the  volumetric  method  developed  in  our 
laboatory. The result shows that neutron  detection  efficiency  of 
Freon-22 for the neutron energy spectrum  obtained  from  an  Am-Be 
source, is almost double, while the life time is 58.6\% smaller than 
that of Freon-12, for a particular neutron flux of that source.
\end{abstract}
\vskip 36 pt

\noindent{\bf 1. INTRODUCTION}
\vskip 24 pt

	A liquid maintained at the same state above its boiling  point 
is said to be superheated. It is a metastable state of  the  liquid 
and can be nucleated to form vapour  by  the  deposition  of  small 
energy  by  ions,  charged  particles  or  by   any   heterogeneous 
nucleation  sites  such  as  gas  pockets,  impurities   etc.   The 
superheated drops,suspended in gel can be used to detect  neutrons  through  the 
nucleation induced by the recoil nuclei in the medium . The  recoil 
nuclei are produced  by  collision  of  neutrons  with  the  nuclei 
constituting the superheated drops.
     The application of superheated drop detector (SDD) in  neutron 
dosimetry has already been established (Apfel {\it et al.},  1984,  1989; 
Ing, 1986) and several  other  potential  applications  of  SDD  in 
neutron research has been discussed  (Apfel,  1979a,  1979b,  1981; 
Chakraborty {\it et  al.},  1990  ).  Apfel  (1992)  has  developed  and 
characterised  a  passive  superheated  drop  dosemeter   using   a 
volumetric technique for neutron monitoring  of  personnel  and  in 
accelerartor applications. These type of dosimeters are also commercially
available from Apfel Enterprises Inc.,USA. Practical application of 
the SDD demands that the sample  must  be  reasonably  stable  against 
spontaneous nucleation due to  background  radiation  and  other environmental 
effects, and at the same time  as  much  as  possible  sensitive  to 
neutrons with energy spectrum  of  interest.  Among  the  different 
important features of the  detector,  the  systematic  quantitative 
evaluation of the sensitivity of SDD  has  been  studied  for  some 
liquids (e.g. Freon-114, Freon-12, Isobutane, Freon 142B)  (Roy  et 
al., 1987) and the response  function  was  reported  (Lo {\it et al.}, 
1988). This paper is  an  attempt  to  investigate  more  sensitive 
liquid for neutron detection. Of the liquids investigated  so  far, 
Freon-12 is considered to be the most sensitive liquid for  neutron 
detection. Since the boiling point of Freon-22 is much  lower  than  
Freon-12, we expect Freon-22 to be more sensitive to neutrons  than 
Freon-12.  To  the  best  knowledge  of  the   authors,   no   such 
investigation has been reported with Freon-22. An  accurate  method 
of determining the nucleation rate  of SDD has  been  developed  by 
Roy, {\it et al.}, (1997b) using a relative manometer. Some other studies 
on the neutron detection sensitivity  and  detector  response  were 
made by Ing and Birnboim (1984), Ing (1986), Ipe {\it et al.}, (1988) and 
Biro {\it et al.}, (1990).
     Nath {\it et al.},(1993) measured the neutron dose equivalent to patients 
undergoing high energy x-ray and electron radiotherapy beams using a SDD device.
They employed a passive method for measuring the total volume of neutron
 induced bubbles by displacing an eqivalent volume of gel into
a graduated pipette. The method for the determination of efficiency of
 detection of neutrons by vapour nucleation of superheated drops using volumetric 
method has been developed in our laboratory by Roy {\it et al.}, (1997a). 
We in this work,  present  the  efficiency  of  neutron  detection, 
maximum nucleation rate and life-time of SDD of Freon-22 irradiated 
by neutrons from an Am-Be source for a particular neutron flux  and 
compare it with those of Freon-12. Some of the physical  parameters 
of  SDD  of  Freon-12  and  Freon-22  are  presented  in   table-1. 
Comparison is made with Freon-12, since Freon-12 is the  most  well 
studied liquid in neutron detection. The paper has  been  organised 
to present a brief outline  of  the  method  of  bubble  formation, 
description of  volumetric  method  and  measurement,  results  and 
discussion.
\vskip 36 pt
\pagebreak

\noindent{\bf 2. THEORY OF BUBBLE FORMATION}
\vskip 24 pt

	The free energy required to form a spherical vapour bubble  of 
radius r in a liquid is given by (Roy et al., 1987)

\begin{equation}
G = 4 {\pi} {r^2}  {\gamma\left(T\right)} - {4\over3} {\pi} {r^3} 
 {\left({p_v} -{p_o}\right)} 
\end{equation}                   
where $\gamma(T)$ is the liquid-vapour interfacial tension,$P_v$ is  vapour 
pressure of the superheated liquid and $P_o$ is the ambient  pressure. 
The difference $p_v$ -$p_o$ is called the {\it degree of superheat} of  a  given 
liquid. One can see from equation (1) that G is maximum at 
\begin{equation}
r = 2 {\gamma\left(T\right)} / {\left({p_v} -{p_o}\right)} = {r_c}                               
\end{equation}
where $r_c$ is called the {\it critical radius}. When a bubble grows to  the 
size of the critical radius it becomes  thermodynamically  unstable 
and grows very fast till the entire liquid droplet vaporises. 
     The minimum amount of energy  (W)  needed  to  form  a  vapour 
bubble of critical size $r_c$ as given by Gibbs (1875) from reversible 
thermodynamics is 
\begin{equation}
W = 16 {\pi} {\gamma{^3}\left(T\right)}/3 {{\left({p_v} -{p_o}\right)}^2}
\end{equation}           

	which is supplied by the energy deposition dE/dx being the energy deposited per unit distance travelled by the nuclei  in  the 
liquid by the recoil nucleus in a path length of  2$r_c$   inside  the 
droplet.
\vskip 36 pt

\noindent{\bf3. PRINCIPLE OF THE VOLUMETRIC METHOD}
\vskip 24 pt

The present method utilizes the superheated drops suspended in 
a dust free viscous elastic gel. The excess  pressure  required  to 
form a vapour bubble of diameter 1mm  inside  this  gel  matrix  is 
usually less than  1mm  of  mercury  (as  observed  in  a  separate 
experiment). Hence the volume of the bubble trapped inside the  gel 
and that at atmospheric pressure are almost equal. This is why  the 
volume change upon nucleation would be the same whether the  bubble 
formed is trapped  inside  the  gel  or  liberated  from  it.  Upon 
nucleation the increase in volume of the droplet would displace the 
trapped air  inside  a  vial  containing  sample.  In  the  present 
volumetric method described in detail  in  the  next  section,  the 
measurement of rate of change of volume has been made using such an 
air displacement system. The superheated drops suspended in the gel 
do not touch each other physically and they nucleate  in  a  random 
manner, independent of each other. Under such conditions it can  be 
shown that the  number  of  the  drops  and  hence  the  volume  of 
superheated liquid would decay exponentially with time. The rate of 
change  of  nucleated  volume  varies  exponentially  with  a  time 
constant $\tau$ (lifetime). The life time of these droplets in  presence 
of a neutron flux and the efficiency of neutron detection has  been 
studied by measuring the volume of vapour formed upon nucleation.

	If neutrons  of flux $\psi$ are incident on superheated drops of volume  V,  liquid 
density $\rho_L$ and molecular weight M  the    vaporization 
rate is given by 

\begin{equation}
\frac{dV}{ dt} = V\psi {\frac{N_A\rho_L}{M}} \eta d \sum{n_i\sigma_i} 
\end{equation}

\begin{tabbing}
\noindent where \= N$_A$ \= = \= Avogadro Number\\
        \> d \> = \> average droplet volume\\
        \> n$_i$ \> = \> number of nuclei of the ith element of the molecule 
whose\\
         \> \>   \> neutron nucleus elastic scattering cross section is 
$\sigma_i$\\
        \> $\eta$ \> = \> efficiency  of  neutron  detection.
\end{tabbing}   

	Due to nucleation by neutrons, the change in position  (h)  of 
the water column along a horizontal glass tube which  is  connected 
to the sample vial (arrangement is given in  next  section),  could 
be measured with respect to time (t). The present set  up  is  made 
completely free from any leakage. So the rate of  increase  of  the 
volume of vapour during nucleation should be equivalent to the rate 
of decrease of the volume of the superheated liquid.  So,  we  have 
this equation     
\begin{equation}
\rho_V A{\frac{dh}{dt}} = -\rho_L{\frac{dV}{dt}}  = -\rho_L 
V\psi{\frac{N_A\rho_L}{M}} \eta d \sum{n_i\sigma_i}
\end{equation}

\begin{tabbing}
\noindent where \= A \= = cross section of the horizontal tube\\
        \> $\rho_V$ \> = density of Freon vapour\\
   \> V \> = volume of superheated liquid at any instant of time t.
\end{tabbing}

        Integrating and solving equations (3)  one can obtain

\begin{equation}
{{h A}\over{m}} = {a\left[1 - \exp{- b\left(t - t_o\right)}\right]}
\end{equation}

\noindent where a = ${{{\rho_L} V_o}\over{\rho_V m}}$  and    
\begin{equation}
b   =   {1\over\tau}    =    \psi{\frac{N_A\rho_L}{M}}    \eta    d 
\sum{n_i\sigma_i}\\
\end{equation}
$hA\over{m}$ is the volume of accumulated vapour in time t per unit 
mass of the sample containing Superheated drops and gel, m  being  the  mass  of 
total sample (gel + superheated drops).  Therefore  the  efficiency  of  neutron 
detection $\eta$ is given by 
\begin{equation}
\eta = \frac{b M}{ \psi N_A \rho_L d  \sum{n_i\sigma_i}}\\
\end{equation}
V$_o$ = initial volume of the Freon drops.
t$_o$ = initial time at which the experiment has been started.
$\tau$ = {\it life-time} of SDD in presence of source.
and the product ab gives the maximum nucleation rate.
The equation (6) has been scaled to per unit mass of the sample for 
standardization in making comparison of two samples.
     Fitting equation (6) for  different  values  of $hA\over{m}$  and  t, 
constants a, b are obtained. Knowing b, the life time $\tau$  (=1/b)  in 
presence of neutron flux  $\psi$  is  obtained.  The  neutron  detection 
efficiency $\eta$ can be found out from equation (8) for a known flux of 
neutrons 
\vskip 36 pt

\noindent{\bf4. EXPERIMENT}
\vskip 24 pt
The experiment was performed with 3Ci  Am-Be  neutron  source, 
which has an energy distribution of neutrons  with  peak  near  3.5 
MeV. The source was placed inside a chamber through which  neutrons 
coming in a fixed range of direction were  used  to  irradiate  the 
SDDs. Two separate sets of experiments were done with Freon-12  and 
Freon-22 respectively at about 30$^o$C and at atmospheric pressure. 
The superheated drops are  usually  suspended in an immiscible gel. 
The gel that we used here, was a  homogeneous mixture of some 
ultrasonic gel and glycerol in suitable proportion. The detail of the 
preparation of the  sample  was  given  elsewhere (Roy et al., 1997b). 
The experimental apparatus consists of long glass tube of cross section 
0.1573 sq.cm, placed horizontally on a graduated platform. The tube 
contained a coloured water column as an indicator of the volume of the 
vapour formed on nucleation. One end of the glass tube was connected 
to the glass vial containing the sample, by means of rubber tube. The 
glass vial and the  horizontal tube were at the same height. So the 
pressure inside and outside the tube were equal, e.g. both were at the 
same  atmospheric  pressure. Therefore the displacement of the water 
column of 1 cm length due to nucleation would be directly related to 
the volume of the vapour formed. These displacements of the water column 
along the glass tube were measured as a function of time.
	The gamma sensitivity of sample of Freon-22 was tested with 
$^{241}$Am (59.54keV), $^{60}$Co (1170keV,1332kev), 
$^{137}$Cs (662keV) gamma sources. The sample was placed either just
in contact or at a close distance to the source. The sensitivity of the sample 
for 4.43MeV gamma ray present in $^{24 1}$Am-Be source has also been tested 
by substantially reducing the neutron dose. 
\vskip 36 pt

\noindent{\bf 5. RESULTS} 
\vskip 24 pt
	The chemical formula and other physical  parameters  including 
critical radius (r$_c$ ) and minimum energy required for nucleation (W) 
as calculated   using  equations  (2)  and  (3)  respectively,  for 
Freon-12  and  Freon-22  have   been   listed   in   Table-1.   The 
experimentally observed variation of volume of vapour formed, $hA\over{m}$ 
(per unit mass of the sample) during nucleation  as a  function  of 
time for Freon-12  and  Freon-22  are  shown  in  Fig.1  and  Fig.2 
respectively. The measured data on rate of nucleation and life-time 
of superheated drops  of  Freon-12  and  Freon-22  in  presence  of 
neutrons from a 3 Ci Am-Be source and  the  efficiency  of  neutron 
detection are given in table-2. 
	The samples irradiated by gamma sources did not show any noticeble
nucleation at the experimental temperature.  

\vskip 36 pt

\noindent{\bf 6. DISCUSSION}
\vskip 24 pt

	From the tabulated results it  is  seen  that  the  degree  of 
superheat (defined by the difference  of  vapour  pressure  of  the 
superheated liquid $p_v$ and the ambient  pressure $p_o$ )  attained  for 
Freon-22 at room temperature (30$^o$C) and at atmospheric pressure  is 
about 38\% larger than that of  Freon-12.  This  indicates  that  in 
presence of neutrons the life time of Freon-22  should  be  smaller 
than that of Freon-12.  Our  experimental  results  show  that  the 
maximum nucleation rate of Freon-22 is about  38\%  larger  and  the 
life time of Freon-22 is about 58.6\% smaller than that of  Freon-12 
for a fixed neutron flux from an Am-Be source. As can be seen  from 
the chemical formula that both of  Freon-12  and  Freon-22  contain 
carbon, chlorine and fluorine while Freon-22 contains one  hydrogen 
replacing one chlorine in  Freon-12.  The  neutron-nucleus  elastic 
scattering cross-section for Freon-12 is 1.89 barns while  that  of 
Freon-22 is 1.69 barns at neutron  energy  3.5  MeV.  Although  the 
probability of interaction of a neutron with nuclei of Freon-12  is 
larger than Freon-22, our experimental results (table-2) show  that 
the present prepared sample of Freon-22 is about twice as efficient 
to detect neutrons than Freon-12. This is due to the fact  that  as 
seen from equation (3), the  energy  required  for  nucleation  (W) 
decreases with degree of superheat and as is evident fron  table-1, 
the saturation vapour pressure of Freon-22 at 30$^o$C is  higher  than 
that of Freon-12,  therefore  at  this  temperature  Freon-22  will 
attain a higher degree of superheat which means a smaller amount of 
energy deposition is required for nucleation than that in Freon-12. 
So, although less number of recoil nuclei are available in Freon-22 
from neutron nuclei elastic scattering, the  percentage  of  nuclei 
capable of deposition of energy greater than W for Freon-22 must be 
larger than that in Freon-12. As a result the efficiency of neutron 
detection of SDD of Freon-22 is larger than that of Freon-12.
	From the experiment with gamma sources it is clear that the SDD
based on Freon-22 is insensitive to gamma at the experimental temperature.
The high efficiency of neutron detection and the insensitivity towards gamma 
makes Freon-22 a suitable superheated drop detector for neutron detection.    

\pagebreak
        
\noindent{\bf REFERENCES}
\vskip 24 pt
\noindent{Apfel R.E. (1979a)  Detector and  dosimeters  for  neutrons  and 
other radiations. {\it US Patent} 4,143,274.}\\
\noindent{Apfel R.E. (1979b) The superheated drop detector. {\it Nucl. Inst. Meth.}
 {\bf 162}, 603.}\\
\noindent{Apfel  R.E.  (1981) Photon-insensitive,thermal  to  fast  neutron 
     detector. {\it Nucl. Inst. Meth.} {\bf 179}, 615.}\\
\noindent{Apfel R.E. (1992) Characterisation of new passive superheated drop (bubble)
     dosemeters. {\it Rad. Prot. Dos. } {\bf 44}, 343.}\\
\noindent{Apfel R.E. and Lo Y.C. (1989) Practical  neutron  dosimetry  with 
     superheated drops. {\it Health Phys.} {\bf 56}, 79.}\\
\noindent{Apfel R.E. and Roy S.C.(1984) Investigation on  the  applicability 
     of superheated drop detector in neutron dosimetry. {\it Nucl. Inst. Meth.} 
{\bf 219}, 582.}\\
\noindent{Biro T.,Kelemen A and Pavlicsek I. (1990) Acoustic  Detection  of 
     neutrons by bubble detectors. {\it Nucl. Tracks Radiat. Meas.} {\bf 17},
 587.}\\
\noindent{Chakraborty K.,Roy P.,Vaijapurkar S.G. and Roy S.C.  (1990) Study 
     on neutron spectrometer using superheated drop detector. {\it Proc. 
of 7th National Conference on Particles and Tracks}, Jodhpur pp 133.}\\
\noindent{Gibbs J.W. (1875) Translations  of  the  Connecticut  Academy  III, 
     p.108.}\\
\noindent{Ing H. (1986) The status of the bubble-damage polymer detector. {\it Nuclear Tracks.} 
{\bf 12}, 49.}\\
\noindent{Ing H. and Birnboim H.C. (1984) A bubble damage polymer detector for 
     neutrons. {\it Nucl. Tracks and Radiat. Meas.} {\bf 8}, 285.}\\
\noindent{Ipe N.E., Busick D.D. and Pollock R.W. (1988)  Factors affecting the 
     response of the bubble detector BD-100 and a comparison of its 
response to CR-39. {\it Rad. Prot. Dos.} {\bf 23}, 135.}\\ 
\noindent{Lo Y.C. and  Apfel  R.E. (1988)  Prediction  and  experimental 
     confirmation of the response function  for  neutron  detection 
using superheated drops. {\it Phys. Rev. A} {\bf 38}, 5260.}\\
\noindent{Nath R, Meigooni A.S., King C.R., Smolen S. and d'Errico F. (1993)
     Superheated drop detector for determination of neutron dose equivalent to
patients undergoing high energy x-ray and electron radiotherapy. {\it Medical
Phys.} {\bf 20}, 781.}\\
\noindent{Roy B, Chatterjee B.K., Das Mala and Roy S.C.  (1997a) Study  on 
     nucleating efficiency of  superheated  droplets  by  neutrons. {\it Radiation Physics and Chemistry} (accepted for publication).}\\
\noindent{Roy B., Chatterjee B.K. and Roy S.C.(1997b)  An accurate  method  of 
     measuring life time of superheated  drops  using  differential 
manometer. {\it Radiation Measurements}  (accepted  for  publication).}\\
\noindent{Roy S.C., Apfel R.E. and Lo Y.C. (1987) Superheated drop detector: 
     A potential tool in neutron  research. {\it Nucl. Inst. Meth.} {\bf A255}, 199-206.}\\

\noindent{\bf Table-1 :} A comparison of the physical parameters of 
Freon-12 and Freon-22
\vskip 24pt
\begin{tabular}{l c c} \hline\\
 & {\bf Freon-12} & {\bf Freon-22}\\
 & &\\ \hline\\
 & &\\
1. ChemicalFormula & CCl$_2$F$_2$ & CHClF$_2$\\
 & &\\
2. Molecular weight & 120.91 & 80.47\\
 & &\\
3. Boiling Point & -29.79 $^o$C & -40.75 $^o$C\\
 & &\\
4. Surface tension ($\gamma$) & &\\
   (at 30$^o$C) dyn/cm & 9 & 8\\
 & &\\
5. Vapour Pressure (p$_v$) & &\\
   dyn / sq. cm. &  7.4556  $\times$  10$^6$  &  1.14777 
$\times$ 10$^7$\\
 & &\\
6. Density ($\rho_L$) gm / cc & 1.293 & 1.175\\
 & &\\
7. Degree of superheat & &\\
 (p$_v$  -  p$_o$)  dyn / sq. cm  &  6.441639$\times$  10$^6$  & 
1.0463739$\times$ 10$^7$\\
 & &\\
8. Critical radius & & \\
${r_c} =  {{2\gamma(T)}\over{\left({p_v}  -  {p_o}\right)}}$  cm  & 
2.79$\times$ 10$^{-6}$ & 1.53$\times$ 10$^{-6}$\\
 & &\\
9. Minimum energy required (W) keV & &\\
 to form a vapour bubble of size r$_c$ & 0.184 & 0.049\\
 (W = ${{16\pi{\gamma^3}(T)}\over{3{({p_v} - {p_o})^2}}}$)   &   &\\ 
 & &\\
\hline\\
\end{tabular}

\pagebreak

\noindent{\bf Table - 2:} Observed results on nucleation \\
\noindent{(Neutron flux = 2.5374$\times$10$^7$ /cm$^2$/s; Peak neutron  energy  =  3.5 
MeV).} 

\vskip 24pt
\begin{tabular}{l c c} \hline\\
& &\\
& {\bf Freon-12} &{\bf Freon-22}\\
& &\\ \hline\\
& &\\
1.Initial Nucleation rate$^1$ (cm$^3$/gm/s)&3.0268 $\times$ 10$^{-4}$  &4.9205 
$\times$ 10$^{-4}$\\
& &\\
2. Lifetime $\tau$ (s) & 3089.59 & 1279.13\\
& &\\
3. Neutron detection efficiency$^2$ & 0.2825$\%$ & 0.5588$\%$\\
& &\\ \hline\\

\end{tabular}

\noindent{{\small 1}  Initial nucleation rate is $\left({1}\over{m}\right)$${dV}\over{dt}$
at the initial time, where m is the mass of the sample and ${dV}\over{dt}$
is the rate of volume change upon nucleation.}\\

\noindent{{\small 2}  neutron detection efficiency is the percentage of neutrons causing nucleation.}\\ 

\pagebreak

\noindent{\bf FIGURE CAPTIONS}

\vskip 24 pt

\noindent{\bf Fig. 1:} Observed variation of volume of vapour formed as a
function of time for Freon-12; ${h A}\over{m}$ = vapour volume per unit mass
of the sample.
\vskip 36 pt

\noindent{\bf Fig. 2:} Observed variation of volume of vapour formed as a
function of time for Freon-22; ${h A}\over{m}$ = vapour volume per unit mass
of the sample.

\end{document}